# FDTD analysis of photonic quasicrystals with different tiling geometries and fabrication by single beam computer-generated holography

G Zito,[1,2] B Piccirillo,[1,2] E Santamato,[1,2] A Marino,[1,3] V Tkachenko[3] and G Abbate[1,3,4]

[1] Dpt. of Physics, University Federico II of Naples, via Cinthia, 80126 Naples, Italy
[2] CNISM, Consorzio Nazionale Interuniversitario per le Scienze fisiche della Materia c/o Physics Dpt. University of Naples Federico II, via Cinthia, 80126 Naples, Italy
[3] CNR-INFM - Coherentia c/o Physics Dpt. University of Naples Federico II, via Cinthia, 80126 Naples, Italy
[4] CNR-INFM Licryl - Liquid Crystal Laboratory c/o Physics Dpt. University of Calabria 87036 Rende (CS), Italy

E-mail: gianluigi.zito@na.infn.it and giancarlo.abbate@na.infn.it

**Abstract**
Multiple-beam holography has been widely used for the realization of photonic quasicrystals with high rotational symmetries not achievable by the conventional periodic crystals. Accurate control of the properties of the interfering beams is necessary to provide photonic band-gap structures. Here we show, by FDTD simulations of the transmission spectra of 8-fold quasiperiodic structures, how the geometric tiling of the structure affects the presence and properties of the photonic band-gap for low refractive index contrasts. Hence, we show an interesting approach to the fabrication of photonic quasicrystals based on the use of a programmable Spatial Light Modulator encoding Computer-Generated Holograms, that permits an accurate control of the writing pattern with almost no limitations in the pattern design. Using this single-beam technique we fabricated quasiperiodic structures with high rotational symmetries and different geometries of the tiling, demonstrating the great versatility of our technique.

## 1. Introduction

Quasicrystals are structures exhibiting long-range aperiodic order and rotational symmetry [1-3]. Mesoscale quasicrystals may possess photonic bandgaps (PBGs) [4-6] that are more isotropic than in conventional photonic and, hence, the PBG becomes more spherical leading to interesting properties of light transmission [7], wave guiding and localization [8], increasing the flexibility of these materials for many photonic applications, also due to the possible presence of many non equivalent defect sites. To construct two-dimensional (2D) or three-dimensional (3D) quasicrystals is a very difficult task. Previously used to realize periodic photonic crystals [9,10], holographic lithography was recently proposed and used to realize quasicrystals too at the mesoscale [11-15]. The holographic lithography is based on the interference pattern of many coherent light beams, that are usually obtained by splitting a single laser beam by suitable grating [11,13], prism [15] or dielectric beam splitters [16], in a single or multiple exposure process [12]. Realizing a quasicrystal structure, exhibiting $N$-fold symmetry [14], requires to control the amplitude and phase of $N$ interfering laser beams, leading to several difficulties. By controlling the relative phases of the interfering beams, different geometries in the tiling of the dielectric medium (typically rods, in a binary pattern, corresponding to the maxima positions of the light distribution) are achievable [15]. Moreover, aperiodic structures cannot be realized even in principle by multiple-beam interference, as for instance the Thue-Morse structure, defined by recursive substitutional sequences [17-20]. One-dimensional and two-dimensional Thue-Morse structures are known to exhibit PGB with interesting omnidirectional reflectance [21-24].

In this communication, we demonstrate the importance of the tiling geometry in the arrangement of a quasicrystal structure. We show, by FDTD (Finite Difference Time Domain) simulations of the transmission spectra in several 8-fold quasiperiodic patterns, the influence of different building tile geometries on the photonic band-gap for low refractive index contrasts. This demonstrates the importance of an accurate control of the writing pattern to produce feasible photonic band-gap structures with low refractive index materials. The typical interference patterns obtained changing the relative phases of eight interfering light beams in a multiple-beam process (examples are shown in [15]) produce structures with important differences in the dielectric distribution with respect to the standard Ammann-Beenker octagonal tiling of space with "squares" and "rhombuses" of equal side lengths $a$ [25,26]. We chose the octagonal case because its building tile is easier to analyze with respect to other structures like Penrose or dodecagonal. We analyze and compare both the structures (octagonal and 8-fold interferential pattern with different unit-tiles) to provide a comprehension of the behavior of the photonic band-gap with respect to the building tile, the variation of the refractive index difference $\Delta n$ and the filling factor (high dielectric constant area to overall area ratio).

Moreover, in this communication, we report the fabrication of several 2D quasicrystals in the mesoscale range, using a single-beam technique based on the spatial modulation of the optical beam by means of Spatial Light Modulator (SLM) and Computer-Generated Hologram (CGH). This method was discussed in details in a previous work [27]. Here we show interesting results from the FDTD simulations about quasiperiodic structures that are difficult to achieve by multiple-beam holography, but achievable with the SLM-CGH technique. That being so, in this work we review the most important aspects of our holographic technique and the results already obtained. In fact, this method permits to control with high accuracy the properties of the required photonic structure achieving the desired dielectric distribution. We were able to produce, with single-beam optical setup, 2D quasiperiodic patterns of rotational symmetry as high as 23-fold, and aperiodic patterns that cannot be realized using $N$-beam interference whatever large may be $N$, as for instance the 2D Thue-Morse pattern. These results demonstrate well the potential of our single-beam technique. Our structures were realized by induced photo-polymerization of liquid crystal-polymer composites. Holographic Polymer Dispersed Liquid Crystals (H-PDLCs) are materials that provide good mechanical and optical properties and can be switched by applying moderate external electric fields [16]. Nevertheless, the SLM-CGH technique might be applied to other kinds of photosensitive materials.

## 2. FDTD analysis, results and discussion

The octagonal structure analyzed in this paper was supposed made of dielectric rods with the Ammann-Beenker tiling of space, in air. The positions of the cylinders of radius $r$ are coincident with the vertices of "squares" and "45° rhombuses" with sides of equal length $a$. This structure presents a complete PBG with a very low threshold value for the refractive index difference ($\Delta n$=0.26) between high and low dielectric materials, whereas the gap width to midgap ratio becomes close to 5% for $\Delta n$=0.45 [28]. Therefore, optoelectronics devices based on the octagonal photonic quasi-crystal (PhQC) promise to be realized in silica, a very common telecommunication optical material, or even in soft materials like polymer. Permitting to record large-area photonic quasicrystals in photosensitive materials and exploiting, typically, such substrates, holographic lithography, therefore, represents an important fabrication technique. The writing pattern of light is usually obtained as multiple-beam interference. In fact, the interference irradiance profiles $I(\mathbf{r})$, according to [11,15]

$$I(\mathbf{r}) = \sum_{m=1}^{N}\sum_{n=1}^{N} A_m A_n^* \exp[i(\mathbf{k}_m - \mathbf{k}_n)\cdot\mathbf{r} + i(\varphi_m - \varphi_n)], \quad (1)$$

where $A_m$, $\mathbf{k}_m$, $\varphi_m$, are the amplitudes, the wave vectors and the initial phases of the interfering beams, respectively, give quasiperiodic distributions of the dielectric material in the recording medium. Depending on the threshold level of the photosensitive matter and the exposure time, the filling factor may have different values. Typically, the maxima positions of the light pattern correspond to the high dielectric regions, that, usually, can be approximated with a structure of dielectric rods in air (or other materials). The number $N$ of the interfering beams determines the order of the rotational symmetry of the quasicrystal pattern [29]. The wave vectors of the interfering beams are given by [15]

$$\mathbf{k}_m = \frac{2\pi n}{\lambda}(\sin(\frac{2\pi m}{N})\sin\theta, \cos(\frac{2\pi m}{N})\sin\theta, \cos\theta), \quad (2)$$

where the $\mathbf{k}_m$, $m$=(1, …, $N$), are oriented at angle $\theta$ with respect to the longitudinal $z$-direction, and are equally distributed along the transverse ($x$, $y$)-plane; $n$ is the average refractive index of the photosensitive mixture, and $\lambda$ is the common wavelength of the beams. By adjusting the parameters in equation (1), different interference patterns can be obtained (figure 1). By changing the relative initial phases of the beams, different spatial arrangements of the rods in air may be realized, depending on the resulting interferential profiles. We compared the octagonal pattern with the 8-fold interferential structures obtained for two different sets of the initial phases. For the first pattern, say, 8-fold(A), the phases were supposed to be all equal, that is $\varphi_1 = \ldots = \varphi_8 = 0$, whereas for the second pattern, say, 8-fold(B), the phases were periodically shifted of $\pi/2$, that is $\varphi_1 = \varphi_5 = 0$, $\varphi_2 = \varphi_4 = \varphi_6 = \varphi_8 = \pi/2$, $\varphi_3 = \varphi_7 = \pi$.

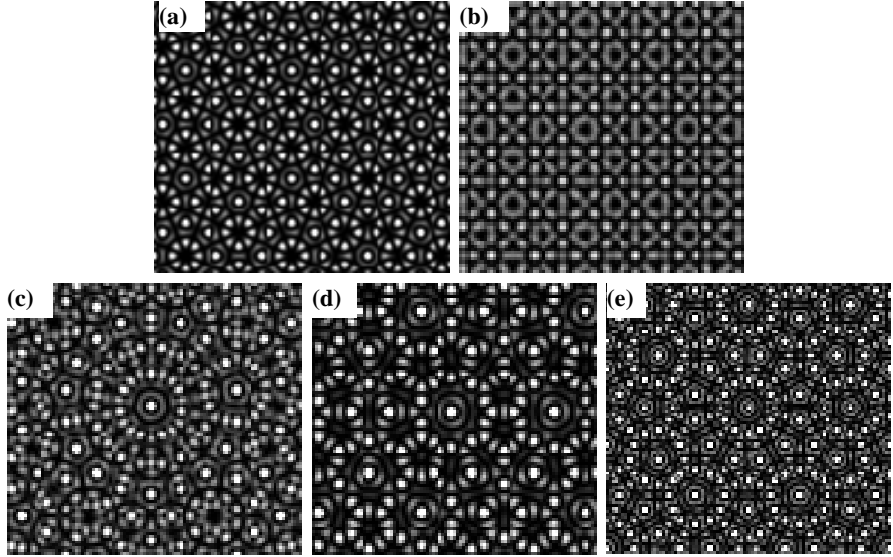

**Figure 1.** (a) Calculated quasiperiodic irradiance profile (IP) with 8-fold rotational symmetry obtained from phase values $\varphi_i = 0$, for $i = \{1,\ldots,8\}$, say, 8-fold(A) pattern; (b) calculated 8-fold IP from phase values $\varphi_1 = \varphi_5 = 0$, $\varphi_2 = \varphi_4 = \varphi_6 = \varphi_8 = \pi/2$, $\varphi_3 = \varphi_7 = \pi$, say, 8-fold(B) pattern. (c)-(e) Calculated IPs from 9-, 10-, 12-beam interference, respectively.

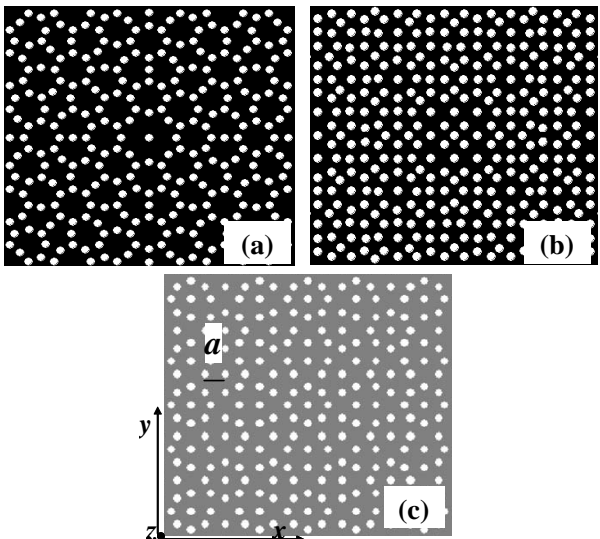

**Figure 2.** (a) dielectric structure of rods from 8-fold(A) pattern; (b) dielectric structure of rods from 8-fold(B) pattern; (c) octagonal structure of rods with "square-rhombus" tile of equal side lengths $a$.

The intensity profiles calculated for both the 8-fold(A) and 8-fold(B) structures are shown in figure 1-(a) and 1-(b), respectively. The corresponding patterns of rods are shown in figure 2-(a) and 2-(b), respectively. These structures are obtained by positioning circular dielectric rods of radius $r$ in the maxima of the interferential intensity pattern IP of figure 1-(a) and 1-(b), respectively. In figure 2-(c), the octagonal structure of rods with the Ammann-Beenker tiling of space is depicted. The side length of the unit-tile of space is $a$ for the octagonal pattern in 2-(c). Due to the non-geometric procedure of building the structures depicted in figures 2-(a) and 2-(b), we found more convenient to define as characteristic length of the patterns a new parameter, that is the average distance $a_d$ between neighbouring rods along the $x$-direction.

The 2D finite difference time domain (FDTD) method with uniaxial perfectly matched layer (PML) boundary conditions was used in all simulations. The photonic quasi-crystals examined were non-periodic in the translational direction and the supercell approximation, necessary in these cases, required very long computational time [30]. The FDTD technique, instead, was faster and very accurate. We employed this approach to obtain transmission information, through the x-y plane (figure 2), as a function of propagation direction, wavelength and polarization. A Gaussian time-pulse excitation was placed in several points of the structure (in different simulations). The pulse was wide enough in frequency domain to cover the range of frequencies of interest. Several detectors were placed in particular positions and allowed us to store field components. Their positions were chosen to cover the angular range related to the 8-fold rotational symmetry (45°) and the mirror symmetry with respect to a line of 22.5° in each 45° sector [31], with an angular separation from 5° to 15°. After a sufficiently large time of calculation, the field was Fourier-transformed to calculate the transmission spectrum with very high frequency resolution. The corresponding wavelength range was (0.1-6.0)μm with a resolution of $\delta=5.0\times10^{-4}$μm. The transmission coefficient through the system was calculated for different values of the dielectric constant of the rods in air, for both polarization TM (electric field $E_z$ parallel to the rod axis) and TE (magnetic field $H_z$ parallel to the rod axis) [5], for each detector, that is for different propagation directions of the time-pulse excitation. The discretization grid provided a minimum of 100 grid points per free space wavelength. The transmission coefficient, normalized with respect to the incident power of the source, was calculated as a function of the filling fraction $r/a_d$, given as the rod radius to average distance ratio. The parameter $r/a_d$ is related to the filling factor and it was varied in order to maximize the gap width of the octagonal structure for fixed refractive index difference $\Delta n=0.65$. We found that interesting values were in the range $r/a_d \approx (0.15 - 0.35)$, with a maximum band width at $r/a_d \approx 0.24$ ($r/a \approx 0.34$). The value of $r/a_d$ was held fixed in all simulations analyzed here for comparison.

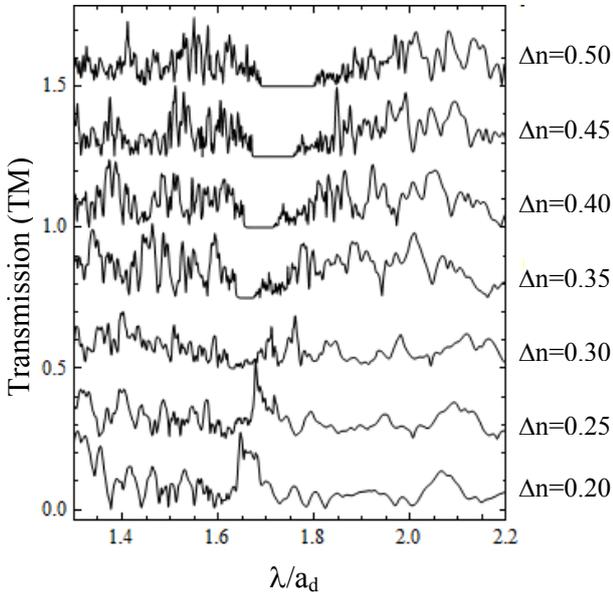

**Figure 3.** Transmission spectra (TM polarization) of the octagonal pattern (Ammann-Beenker tiling) for several values of the refractive index difference $\Delta n$.

The data collected from the detectors placed at different positions and angular orientations had the same overall shape for all the transmission spectra we investigated here, demonstrating the isotropy of the structures with respect to the propagation direction of the excitation source and, hence, the existence of a complete photonic band-gap.

In figure 3, the transmission spectra of the octagonal pattern of figure 2-(c) is shown for TM polarization as a function of the refractive index difference. The spectra are shifted in the vertical direction to permit comparison. We see that an index difference of $\Delta n \approx 0.3$ is needed to form the band-gap at the free space mid-gap wavelength $\lambda_m=1.62$μm (corresponding to the tile length $a=0.93$μm and the average distance $a_d=1.0$μm), with a gap width to midgap ratio $\Delta\lambda/\lambda_m$ of 0.3%. The PBG had $\Delta\lambda/\lambda_m=6.3\%$ at $\lambda_m=1.74$μm for $\Delta n=0.5$ and increased up to $\Delta\lambda/\lambda_m=22.6\%$ at $\lambda_m=2.21$μm for $\Delta n=1.0$ (not shown in figure 3). The mid-gap free space wavelength $\lambda_m$ varied with $\Delta n$ because the increasing of the index difference corresponds to increasing of the average refractive index.

We did not find a clear PBG in the octagonal pattern of figure 2-(c) for the TE polarization. Our simulations showed, in fact, only a very narrow PBG with a relative width <1% for $\Delta n=0.9$ around 1.48μm, whereas in other regions of the spectra the possible presence of a band-gap was completely hidden by structures of multiple peaks that were related, probably, to localized modes (not shown here). Moreover, in the case of the TE polarization, the data collected from the detectors placed at different angular orientations demonstrated a dependence from the propagation direction.

In figure 4-(a) and 4-(b), respectively, the transmission coefficients in TM polarization of the 8-fold(A) and 8-fold(B) patterns (see figure 2a and 2b) are shown, for a refractive index difference $\Delta n=1.5$. The two structures presented important differences. The pattern (A) had a PBG of $\Delta\lambda/\lambda_m=19\%$ around $\lambda_m=3.75$μm, whereas the pattern (B) had two band-gaps, one with a relative width $\Delta\lambda/\lambda_m=13\%$ at $\lambda_m=1.93$μm and the other of 12% at 3.30μm. We analyzed the same structures for several values of the index contrast and TM polarization. We found that, in comparison with the octagonal structure, an index contrast larger of about a factor two was needed to open a PBG of comparable width to mid-gap ratio. The pattern (A) had a larger band-gap, but for longer wavelengths ($\lambda_m=3.75$μm) with respect to the first band-gap of the pattern (B) and with respect to the PBG of the octagonal pattern of figure 2-(c), that had $\lambda_m=2.6$μm for $\Delta n=1.5$ (not shown in figure 3). On the other hand, the 8-fold(B) pattern, even with a smaller PBG, could work at shorter wavelengths ($\lambda_m=1.93$μm). By changing the filling fraction the transmission spectra preserved the overall shape.

Only a shift in the mid-gap wavelength was observed.

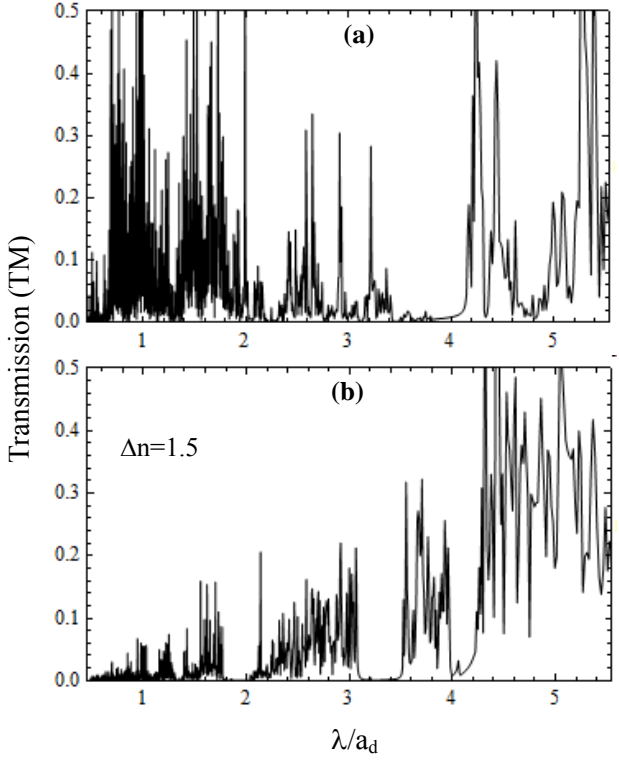

**Figure 4.** (a) Transmission spectrum (TM polarization) for the 8-fold(A) pattern with refractive index difference $\Delta n$=1.5; (b) Transmission spectrum (TM polarization) for the 8-fold(B) pattern with $\Delta n$=1.5.

Very interesting are the results related to the calculation of the transmission spectra for the TE polarization. In figure 5-(a) and 5-(b), the transmission information in TE polarization related to the 8-fold(B) pattern, calculated supposing a refractive index difference $\Delta n$=0.6 and $\Delta n$=0.4, respectively, are shown. Two PBGs, independent from the propagation direction, were present, one for $\Delta n$=0.4 with a relative width $\Delta\lambda/\lambda_m$=32% at the mid-gap free space wavelength $\lambda_m$=1.25µm and the other for $\Delta n$=0.6 with a relative width $\Delta\lambda/\lambda_m$=31% at $\lambda_m$=1.30µm: very short wavelengths in both cases compared to the other patterns. The peaks around 1.3µm were probably due the existence of localized modes. With the increasing of the refractive index difference only a shift in the mid-gap wavelength was observed, up to 2.25µm for $\Delta n$=1.6. The 8-fold(A) pattern, on the other hand, did not present low index PBG for TE polarization (not shown here) as is also for the octagonal structure with the Ammann-Beenker tiling.

The behaviour of the 8-fold(B) pattern represents, in our opinion, a surprising result promising the implementation of reliable low index contrast PhQC devices in polymeric substrates.

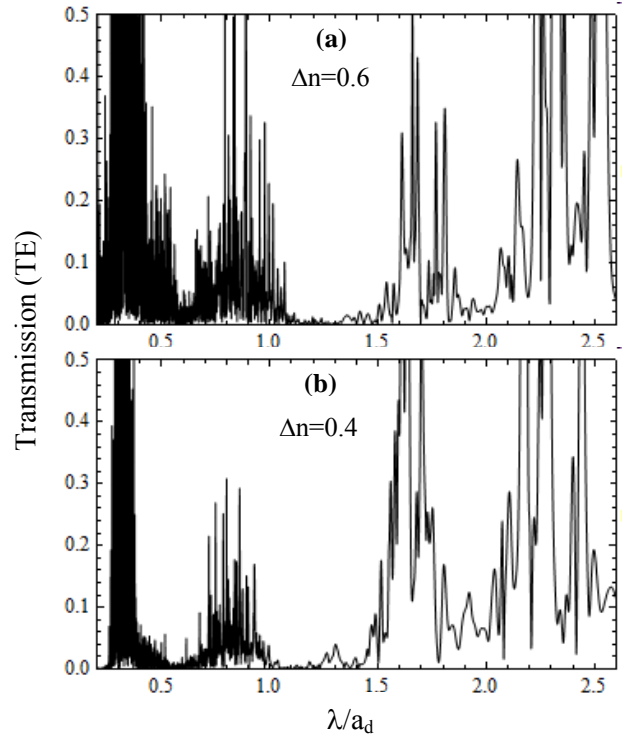

**Figure 5.** (a) Transmission spectrum (TE polarization) for the 8-fold(B) pattern with refractive index difference $\Delta n$=0.6; (b) Transmission spectrum (TE polarization) for the 8-fold(B) pattern with $\Delta n$=0.4.

We analyzed also the complementary dielectric structures of the 8-fold(A) and (B) patterns, that is circular rods of the low index material (air) embedded in the high index substrate. Also in this case no band-gap was observable for the low refractive index contrasts examined. Nevertheless, the simulations were performed only for TM polarization in this case.

## 3. Holographic patterning and fabrication

As demonstrated in the previous section, accurate control of the dielectric arrangement is fundamental to provide the desired low index PBG in quasicrystal structures. The interferential structure designated as 8-fold(B) might be a good candidate for this purpose. For this reason, in this section, we review the Computer-Generated Holography technique, a single-beam technique developed in a previous work [27] that permits to fabricate almost any kind of structure, interference-based or not, with high reproducibility and very accurate control of the light intensity distribution in the writing process.

Computer-generated holography is an attractive technique that allows to create, with almost any design, two-dimensional, even three-dimensional, spatial distributions of the optical beam intensity by controlling the phase profile of a laser impinging on a Diffractive Optical Element (DOE) [32,33]. In our work, we implemented the DOE by means of a programmable liquid crystal SLM that can encode the CGH in its LC-display. While in the standard multi-beam holography is quite difficult to work with a large number of beams [29], with the CGH-SLM it is possible to encode the desired spatial distribution in a single beam. Although limited from the pixel resolution of the SLM, our CGH-SLM technique alleviates the great difficulty in maintaining phase coherence of the interfering beams between subsequent exposures, providing high reproducibility of the resulting structures. Moreover, the real-time switching from one pattern to the other is easily accomplished without changing the optical alignment. We adopted a liquid crystal spatial light modulator HoloEye LC-R 3000, permitting intensity images of 256 grey levels with a maximum resolution of 1200×1920 pixels. The CGHs corresponding to the optical phase profiles to be added to the incident beam have been addressed to the SLM via computer. By using an iterative algorithm it is possible to create the desired irradiance pattern in the Fourier plane of the CGH [32,34,35]. Alternatively, we encoded the desired intensity pattern directly into a phase-only CGH [36], with a "direct imaging" method, discussed in details in [27].

The spatial light modulator LC-R 3000 (for visible light at λ=532nm) had a pixel pitch of 9.5μm. We used a mixture of the monomer dipentaerythrol-hydroxyl-penta-acrylate DPHPA (60.0% w/w), the liquid crystal BLO38 by Merck (30.0% w/w), the cross-linking stabilizer monomer N-vinylpyrrolidinone (9.2% w/w) and a mixture of the photoinitiator Rose Bengal (0.3% w/w) and the co-initiator N-phenylglycine (0.5% w/w) [34,37,38]. The polymer had a refractive index $n_p$=1.530, whereas the LC BLO38 had an ordinary refractive index $n_o$=1.527 and an extraordinary refractive index $n_e$=1.799. The average refractive index was estimated to be n~1.57. Although the refractive index contrast achievable with such a mixture, usually, is Δn ~ 0.2 - 0.3, the liquid crystal content can be removed to increase the index difference up to ~ 0.5. Different mixtures or materials can be used in order to increase the amplitude of the dielectric modulation.

Here we review the results obtained with our technique [27]. We realized quasi-periodic patterns with 8-fold rotational symmetry, in particular the 8-fold(A) and 8-fold(B) pattern described in the previous section (see figure 1a-2a and 1b-2b, respectively). Furthermore, we were able to realize other two-dimensional quasicrystal structures with 9-, 10-, 12-, 17-, 23-fold rotational symmetry, and the aperiodic structure based on the Thue-Morse sequence [17]. Figures from 6-(a) to 6-(h) refer to these structures [39]. Three insets are shown in each figure, consisting of: the calculated Irradiance Profile (IP) sent to our SLM and then imaged at the sample position; the calculated 2D Fourier Transform (FT) of this irradiance pattern; and the experimental Diffraction Pattern (DP) produced by the written structure. These structures are two-dimensional phase gratings in which the modulation profile of the average refractive index is not easily accessible with the scanning electron microscopy. The similarity between the positions and cone angle of the spots in the calculated Fourier transform FT and the observed DP doubtless substantiates the presence of the index profile and the good quality of the samples. The DPs show the expected N-fold symmetry having N points for even symmetry and 2N points for odd symmetry. We estimated the size d of the "self-similarity cell" (that is the self-repeating basic structure observable in the calculated intensity profiles) from the magnitudes of the basic reciprocal vectors in the diffraction patterns, that are related to sensible lengths of the crystal structures, such as the tile side [16,40]. Although the size d reported in figure 6 (IP insets) was in the range between 4.8 and 8.6μm, increasing with the order of the symmetry, the typical separation between equal dielectric regions in the fine structure of the self-similarity cell was in the range of ~1 - 2μm. In fact, we were able to realize 1D Bragg grating and 2D periodic square lattice with a pitch of ~1μm (not shown here). The scale length of the realized structures depended on the lateral magnification of the relay lenses [27] (which also affected the extension of the writing area). The achievable resolution, lastly, depends on the SLM pixel size and on the wavelength of the writing light. Using UV light and state-of-the-art SLM pixel size [41], the limit in the spatial resolution could be improved by a factor of 3-4. Anyway, the scalability of the optical properties permits to use our quasiperiodic structures to study their optical behavior. Furthermore, according to the TE transmission spectra for the 8-fold(B) pattern of figure 5 presented in the previous section, an average distance $a_d$~1.25μm between neighbouring rods would be sufficient to induce a PBG at $λ_m$=1.55μm. That being so, the SLM-CGH technique could be employed to produce photonic quasicrystals for feasible and reliable applications.

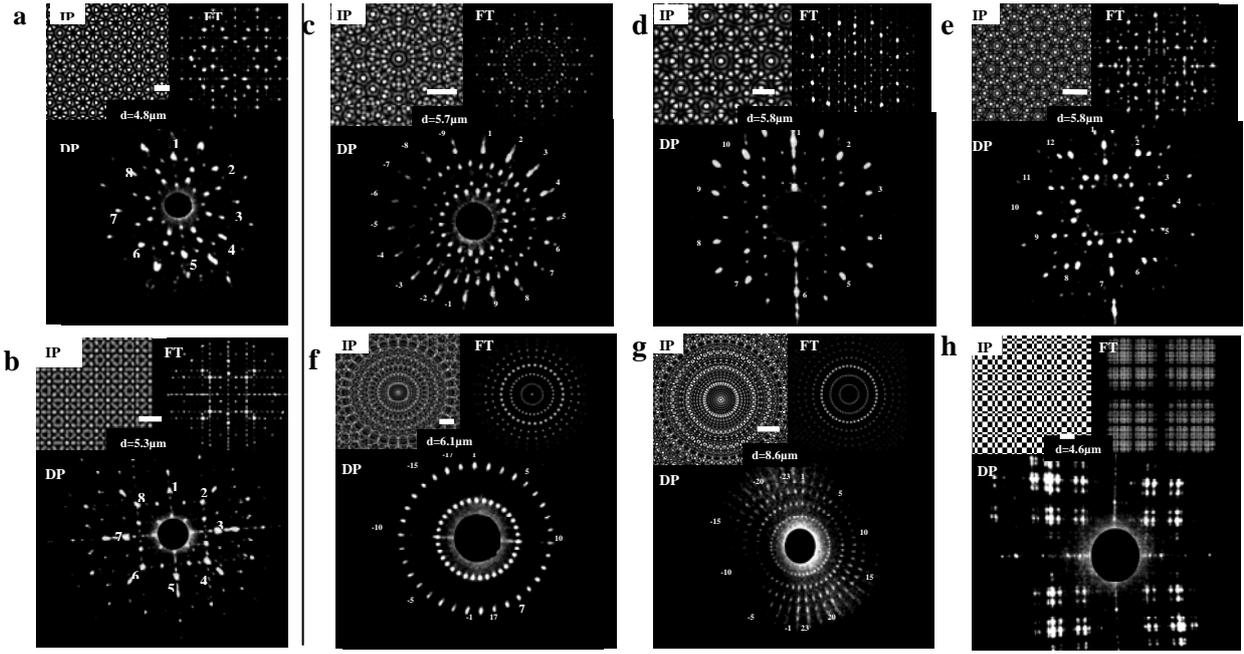

**Figure 6.** (a) Quasiperiodic structure 8-fold(A) with phases $\varphi_i = 0$, for $i = \{1,\ldots,8\}$; (b) quasiperiodic structure 8-fold(B) with phases $\varphi_1 = \varphi_5 = 0$, $\varphi_2 = \varphi_4 = \varphi_6 = \varphi_8 = \pi/2$, $\varphi_3 = \varphi_7 = \pi$. (c)-(g) Quasiperiodic structure with 9-, 10-, 12-, 17-, 23-fold rotational symmetry, respectively. (h) Two-dimensional Thue-Morse quasicrystal structure. (a)-(h) Top left inset: calculated irradiance profile (IP); top right inset: 2D Fourier transform (FT) of the irradiance profile; bottom inset: observed diffraction pattern (DP); $d$ estimates the self-similarity cell size of the structures we fabricated.

## 4. Conclusion and discussion

In conclusion, we showed, by FDTD simulations of the transmission spectra of 8-fold quasicrystals, the influence of different building tile geometries on the photonic band-gap, demonstrating the importance of an accurate control of the writing pattern to produce feasible photonic band-gap structures with low refractive index materials. Besides the octagonal pattern with the Ammann-Beenker geometric tiling, the 8-fold(B) pattern of figure 2-(b) was found very interesting among the interferential structures because it permits to obtain a large PBG for an index contrast as low as 0.4, as shown in figure 5. For this purpose we presented a fabrication technique that alleviates many drawbacks of the multi-beam holographic lithography that is typically employed for the PhQCs realization. We used Computer-Generated Holograms to drive a liquid crystal Spatial Light Modulator. This permits to write arbitrary structures with simple and reliable optical setup. The writing intensity pattern can be modified by real time change of the CGH addressed to the programmable SLM. Accurate control of the pattern designs and of the tiling geometry can be achieved without mechanical motion and optical realignment. Together with the 8-fold quasicrystal patterns discussed in the section 2, we were able to obtain, with a single-beam technique, quasiperiodic structures with unprecedented rotational symmetries up to 17- and 23-fold and two-dimensional Thue-Morse structures too. The possibility of a full characterization of quite interesting aperiodic structures not achievable by multiple beam holography even in principle (like two-dimensional Thue-Morse), and that have been studied until now only theoretically [24], is open. Our structures were written into polymeric liquid crystal films, so to permit switching by external fields [16]. However, our holographic technique could be applied, in principle, to any photosensitive material (e.g. with a larger index contrast), or to produce patterned masks and templates for use in lithography of hard materials [9-16,29]. We expect that the SLM based CGH technique may have a large impact on the production of complex photonic structures.

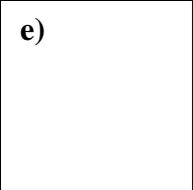
e)